# Preliminary Baseline Antenna Design for the Black Hole Explorer (BHEX) Mission

Robert Lehmensiek∗,†, Sridharan Tirupati Kumara∗
∗ National Radio Astronomy Observatory, Charlottesville, VA, USA
† Dept. of Electrical and Electronic Engineering, Stellenbosch University, Stellenbosch, South Africa

*Abstract*—The Black Hole Explorer (BHEX) mission extends the submillimeter Very-Long-Baseline Interferometry (VLBI) to space. The preliminary baseline design of a shaped axial-symmetric displaced-axis dual-reflector antenna for the BHEX is presented. The main goal of the antenna design optimization is to maximize aperture efficiency given the geometric and mechanical constraints of a space-borne antenna.

## I. INTRODUCTION

THE aim of the Black Hole Explorer (BHEX) mission is to discover and measure the photon ring that is predicted to exist in images of black holes produced from light that has orbited the black hole before escaping. Achieving the necessary angular resolution demands interferometer baselines significantly longer than the diameter of Earth. Thus, the purpose of the BHEX is to extend millimeter/submillimeter Very-Long-Baseline Interferometry (VLBI) to space [1]. To meet the scientific objectives, the BHEX mission requires high-sensitivity observations. This places rigorous constraints on the antenna's performance, as well as its size, weight, and power.

In this paper we present the preliminary baseline design of a large, lightweight, space-borne antenna that can operate at millimeter/sub-millimeter wavelengths with maximized aperture efficiency. The mission background and the antenna technology study are discussed in an accompanying paper [2]. An overview of the mission instruments and the conceptual design of the antenna which has matured into the current baseline design were previously presented [3, 4, 5].

A circular aperture with diameter of 3.4 m is specified to meet both the system's gain specification and manufacturability. The main performance metric for this space-borne antenna is its aperture efficiency, $\eta_a$, unlike typical earth-based telescopes that strive to maximize the effective aperture area over the system temperature, $A_e/T_{sys}$ [6]. The BHEX instrument system operates in two frequency bands: 75 GHz – 107 GHz, and 225 GHz – 320 GHz. All performance parameters presented in this paper were determined by Physical Optics (PO) augmented by the Physical Theory of Diffraction (PTD) technique as implemented in GRASP [7].

## II. CLASSICAL AXIALLY SYMMETRIC REFLECTOR SYSTEMS

An axially symmetric reflector system is the preferred choice from a mechanical stiffness perspective, specifically considering the antennas size and large mechanical loads during launch. Dual reflectors with their folded optics give a compact structure and allow for the feed to be mounted behind the primary reflector out of the way of the optical path. Classical dual reflectors have either a Cassegrain or a Gregorian configuration, where the sub-reflector is either a hyperboloid or ellipsoid shape respectively, and the primary is a paraboloid. Due to the sub-reflector in the optical path, these systems have central blockage which can cause feed mismatch and gain ripple over frequency. The preferred choice then is the displaced axis version. In this case the primary reflector's focal axis is displaced from the axis of symmetry as shown in Fig. 1 for a classical Gregorian system which has a real displaced focus off-axis and the reflector system's focal point on the symmetry axis. The offset Cassegrain system has a virtual focus behind the sub-reflector. The main advantage of the offset configuration is the clear optical path (ignoring the sub-reflector's struts). The sub-reflector's central blockage reduces the total aperture size slightly, but does not cause feed mismatch.

For the offset configuration, depending on how the axis is offset, four possible dish geometries exist [8, 9]. However, two of these can cause sub-reflector self-blockage and will thus not be considered here. The two remaining classical configurations are the axially displaced Cassegrain (ADC) and the axially displaced ellipse (ADE), adopting the naming convention as in [9]. These dishes are completely defined by the following set of parameters (also see Fig. 1):

- $D_m$ – diameter of the primary reflector
- $D_s$ – diameter of the sub-reflector
- $F$ – focal distance of the primary reflector
- $\theta_e$ – feed subtended half-angle
- $D_b$ – diameter of blockage

For the BHEX antenna $D_m$ is given as 3.4 m, and we chose $D_b$ equal to $D_s$ to minimize blockage. For the remaining three parameters we have performed an exhaustive parametric study using Gaussian feed patterns with various edge tapers as input to the reflector system. The chosen parameter space was as follows:

- $D_s / D_m \in [0.05, 0.15]$
- $F / D_m \in [0.2, 0.41]$
- $\theta_e \in [2°, 12°]$

The results of this parametric study showed that the maximum aperture efficiency in the case of the classical ADC system is around 82% with a Gaussian feed pattern that has an edge taper of -11.8 dB. For the classical ADE the maximum aperture efficiency is 91% with a -21 dB edge taper for the Gaussian feed. The reason for the difference in performance will be explained in the next section.

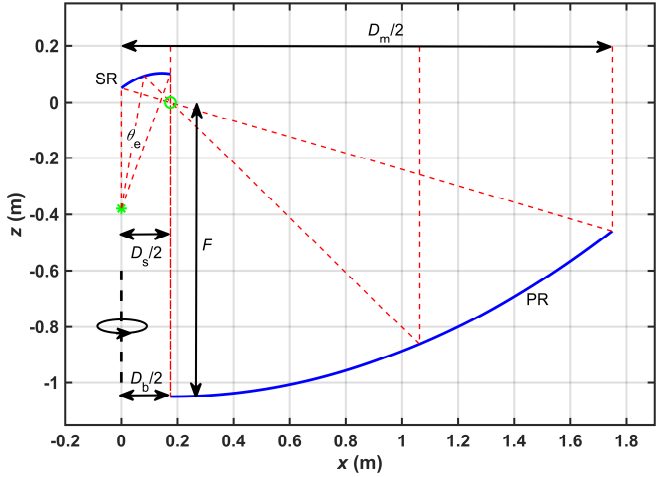

**Fig. 1.** An illustrative classical axial-symmetric displaced-axis dual-reflector system in the Gregorian configuration.

For the sub-reflector size, the trade-off is on the one hand the minimum electrical size required at the lowest frequency and on the other hand the minimization of blockage. The optimal size for a Cassegrain system is in the range 10-15% of $D_m$, and for the ellipse system in the range of 5-10% for our antenna under consideration with $D_m$ equal to 3.4 m.

In general, the dish's aperture efficiency performance is independent of the parameters $F/D_m$ and $\theta_e$, if $\theta_e$ is larger than 6°. A smaller $F/D_m$ is preferred from a surface tolerance loss viewpoint. The impact of these parameters on performance will be discussed in more detail in the next sections.

Besides the aperture efficiency performance metric, a few geometric constraints also need to be adhered to. These are: the maximum height and curvature constraints for manufacturing the primary reflector; sub-reflector position relative to primary vertex to minimize strut lengths and thermal considerations; and the position of the downstream optics to be below the vertex of the primary.

### III. SHAPING FOR MAXIMUM EFFICIENCY

From a Geometric Optics (GO) perspective, a reflector system transforms the primary (feed) radiation field pattern, $G(\theta_f)$ to an aperture plane electric field distribution, $E(\rho)$, and this mapping is defined by a mapping function

$$\rho(\theta_f) = \frac{|G(\theta_f)|^2 \sin\theta_f}{V_c |E(\rho)|^2 \rho'(\theta_f)}. \quad (1)$$

Here $\theta_f$ is the angle measured from the feed axis and $\rho$ is the distance from the projected primary reflector aperture axis, $\rho'(\theta_f)$ denotes the $\theta_f$ derivative of the function $\rho$, and $V_c$ is a normalization constant.

For a reflector system with desired maximized aperture efficiency, we can assume a uniform aperture field distribution. Then, for a given feed pattern, we can determine using (1) the required mapping function. Given the mapping function, the shaping procedure determines the reflector surfaces by GO to trace the rays between the feed point and the aperture plane, enforcing Snell's law of reflection at the surfaces, conservation of power within the ray tubes, and maintaining a constant path length condition for all the rays [10]. As a starting point for the shaping procedure we require a set of three points. These are: the reflector system's focal point, and initial points on the sub- and primary reflectors. These are typically, as is done here, derived from the classic conic-section geometries.

For the classic (unshaped) systems the mapping function and resultant aperture field distributions are shown in Fig. 2(a) and Fig. 2(b) for the ADC and ADE cases respectively. The input pattern (arbitrarily) chosen here was a Gaussian shape with -15 dB edge taper. In both cases the resultant aperture field distribution is not uniform. However, for the ADE case it is much closer to a uniform distribution at least over a large part of the aperture plane. This explains the better performance in aperture efficiency mentioned in Section II. Note that for the ADE geometry the rays from the focus to the aperture plane are reversed, i.e. the central part of the feed's radiation (most power) is mapped to the outside part of the aperture plane (larger area) and vice versa.

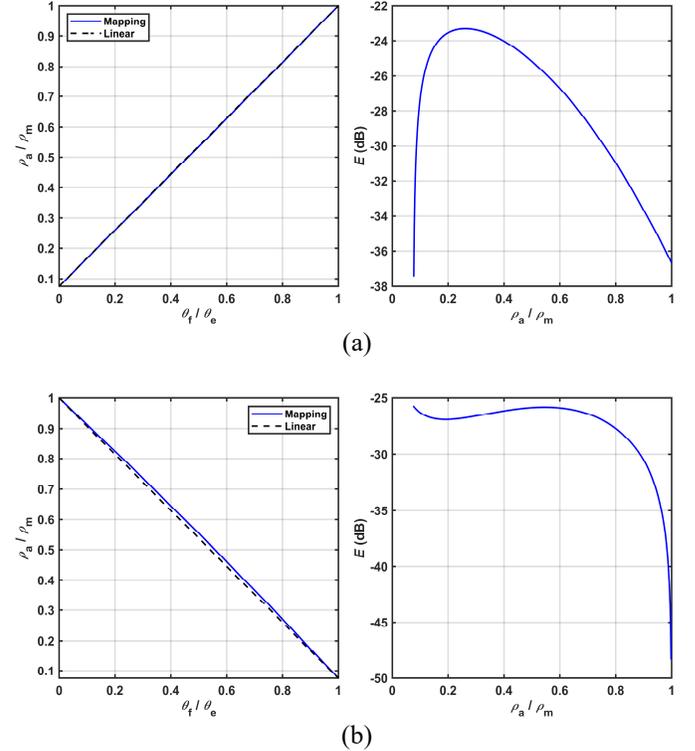

**Fig. 2.** Mapping functions and resultant aperture field distributions for a classical (a) ADC and (b) ADE configuration with a -15 dB Gaussian feed.

When we shape the reflector surfaces for a uniform aperture field distribution, and again assuming a -15 dB edge taper Gaussian feed as input, the resultant mapping is shown in Fig. 3(a) and Fig. 3(b) starting either with the ADC or ADE geometries respectively.

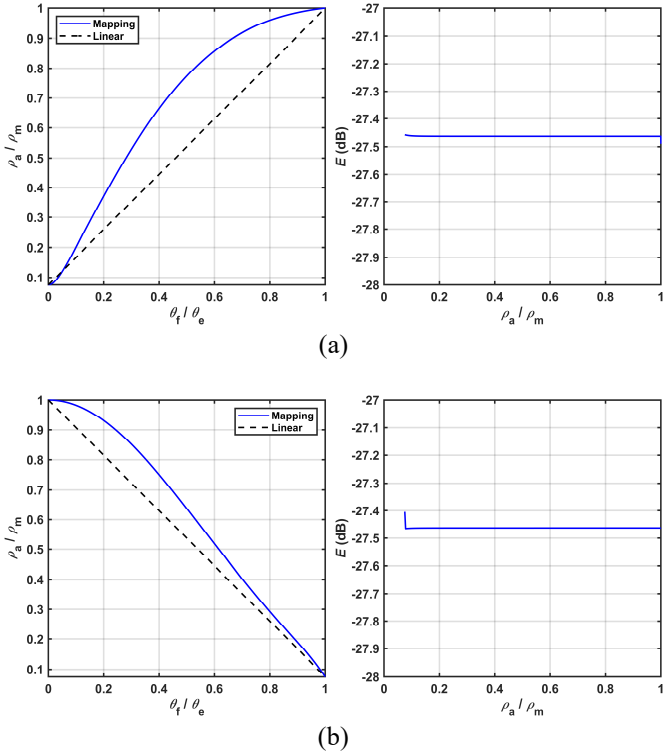

**Fig. 3.** Mapping functions and resultant aperture field distributions for a shaped (a) ADC and (b) ADE configuration with a -15 dB Gaussian feed.

As was done for the classical geometries, we did a full parametric study of the shaped systems using the same parameter space as in Section II. For both systems we achieved an aperture efficiency larger than 95% with a Gaussian edge taper of -20 dB. The higher the edge taper the better as more power from the feed is intercepted. The optimal sub-reflector size was in the range 7-10% of $D_m$. As a smaller sub-reflector size is preferred from a mechanical perspective, the choice was to go forward with $D_s = 0.075\ D_m$. Given that we shape the dish geometry for maximized performance for each parameter set, the efficiency is flat over $F/D_m$ as shown in Fig. 4 at the mid-band frequencies. The aperture efficiency decreases for $\theta_e$ smaller than ~6° at the lower frequencies. Also plotted on the contour plots in Fig. 4 is the surface loss efficiency due to a 40 μm primary and a 10 μm sub-reflector rms surface tolerance [11] which shows that a smaller $F/D_m$ is preferred.

Although both the shaped ADC and ADE systems have similar results, the ADC is preferred due to the convex shape of the sub-reflector which is easier to machine, and the distance between sub- and primary reflectors is slightly smaller. This leads to a slightly more compact structure and easier thermal management (see Section IV).

In summary, from the exhaustive parametric study, a shaped ADC with the following parameter set was chosen for the BHEX antenna:

- $D_m = 3.4$ m
- $D_s = D_b = 255$ mm
- $F/D_m = 0.25$
- $\theta_e = 6.3521°$

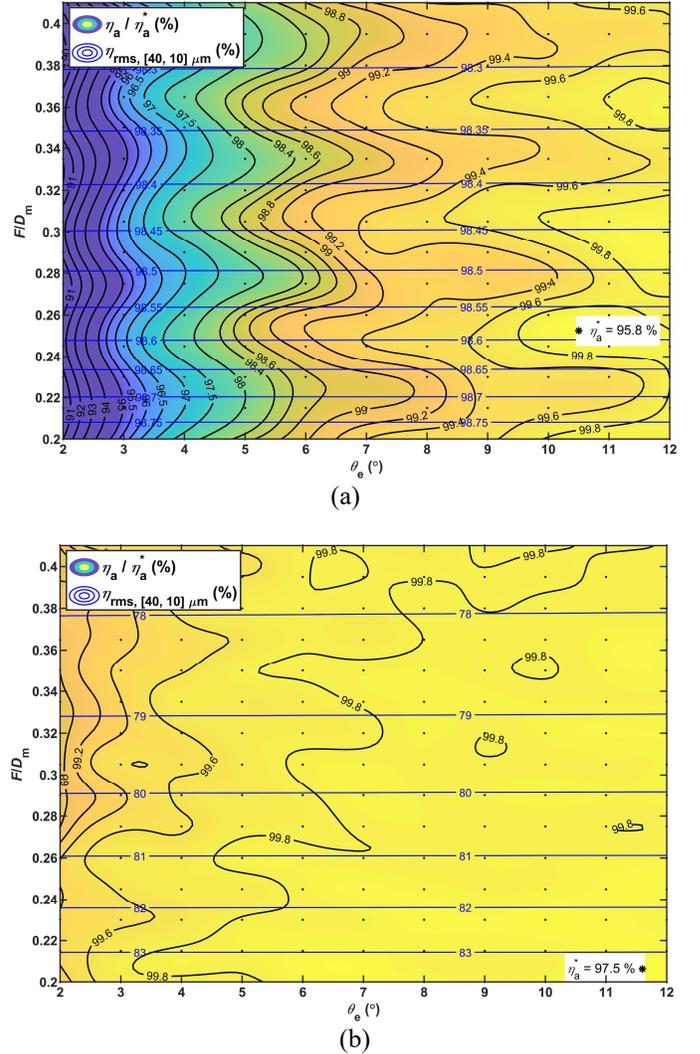

**Fig. 4.** Contour plot of the degradation in aperture efficiency over the parameter space $F/D_m$ and $\theta_e$ for the shaped ADC system with $D_m = 3.4$ m, $D_s/D_m = 0.075$, Feed's Gaussian edge taper = -20 dB at (a) 84 GHz and (b) 320 GHz. The maximum (undegraded) aperture efficiency is indicated with an asterisk. Superimposed on the plot is the surface loss efficiency in percentage.

The geometry is shown in Fig. 5. The depth of the primary in this design is 0.874 m. The feed subtended half-angle was chosen such that the distance from focus to primary vertex is 0.3 m, as required by the downstream optics design. The specification for the downstream optics then is a radiation pattern that is down to at least -20 dB at $\theta_e$ and needs to be axial-symmetric and frequency-invariant. The low edge taper is advantageous for an optimum shaped system, as most of the horn's radiated energy can then be coupled to the aperture plane.

## IV. THE FINAL SHAPED DESIGN

The downstream optics consists of two radially corrugated horns and a set of focusing mirrors. The conceptual downstream optics design [12] meets the specifications of Section III and has a radiation pattern as shown in Fig. 6. The pattern is fairly symmetric and invariant over frequency, apart from the low frequency end.

The mean over frequency and $\varphi$ of this radiation pattern is taken as the input to the shaping algorithm to determine the final shaping. The edge taper is about -24 dB. The resultant reflector system radiation pattern is shown in Fig. 7 and the achieved aperture efficiencies are summarized in Table I (excludes surface tolerance, ohmic and strut blockage losses).

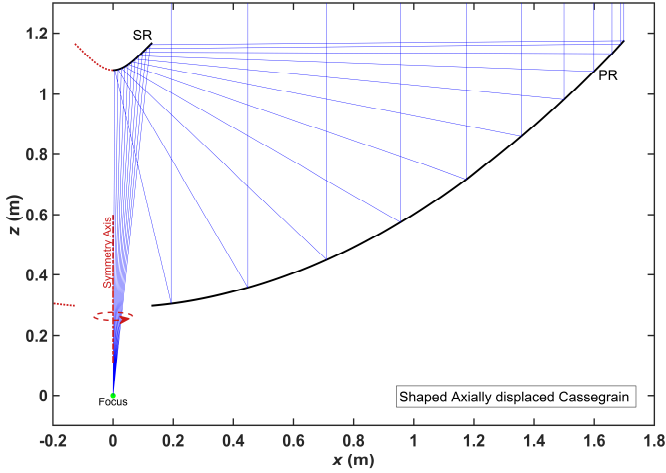

**Fig. 5.** The baseline geometry for the BHEX shaped ADC reflector system.

TABLE I
THE APERTURE EFFICIENCY OVER FREQUENCY

|  | Frequency (GHz) | $\eta_a$ (%) |
|---|---|---|
| Band 1 | 75 | 89.2 |
|  | 83 | 94.9 |
|  | 91 | 95.7 |
|  | 99 | 95.8 |
|  | 107 | 95.9 |
| Band 2 | 225 | 95.9 |
|  | 249 | 96.3 |
|  | 273 | 97.4 |
|  | 296 | 97.5 |
|  | 320 | 97.2 |

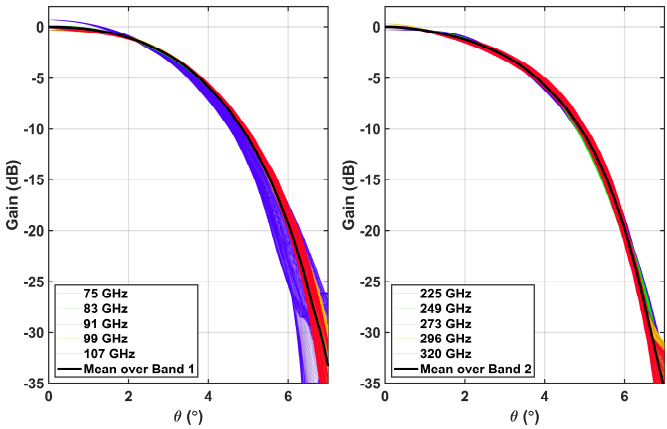

**Fig. 6.** Downstream optics radiation pattern over the extent of the sub-reflector shown at five frequencies in each band and at every 1° cut in $\varphi$.

One advantage of the current dish design is that the sub-reflector is below the outer rim of the primary. This avoids direct sun incidence even for sun angles of 90° from the antenna boresight. Given the small size of the sub-reflector it can be machined to high accuracy from aluminium and it can be temperature controlled.

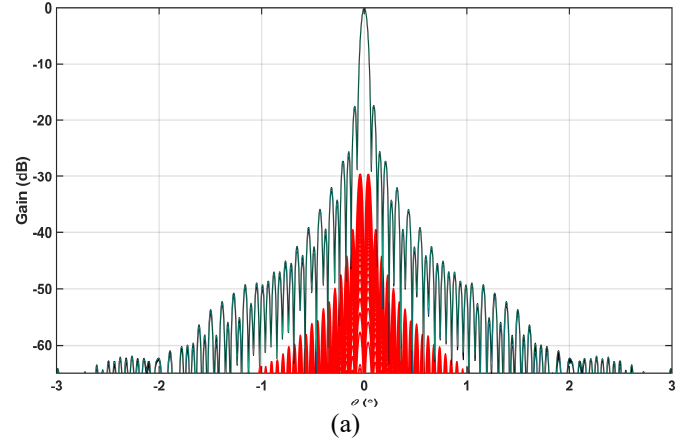

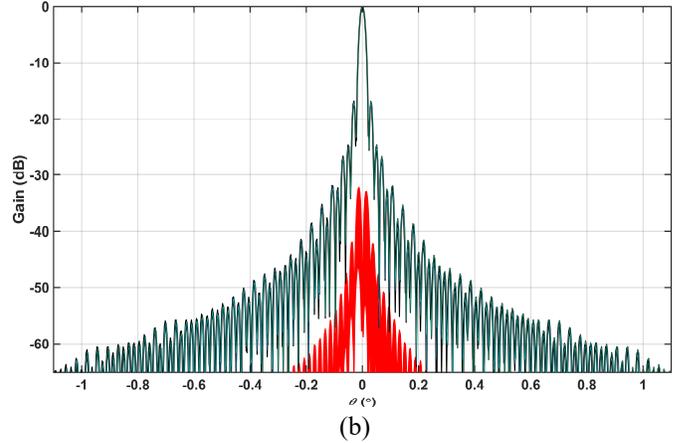

**Fig. 7.** The simulated radiation pattern of the shaped ADC reflector system with the downstream optics radiation pattern as feed at (a) 91 GHz and (b) 273 GHz.

## V. TOLERANCE ANALYSIS

A tolerance study was performed where the feed or the sub-reflector were translated or rotated about their local coordinate frames. This is done for the worst case, i.e. at 320 GHz. The results are given in Fig. 8 for the perturbed feed and in Fig. 9 for the perturbed sub-reflector. Here the feed refers to the whole downstream optics package.

## VI. SUPPORT STRUT CONFIGURATION TRADE

The sub-reflector is held in place by struts. Two options were considered: the tripod and hexapod iso-static configurations. The impact that these struts have on the aperture efficiency is shown in Fig. 10 and Fig. 11. The tripod structure outperforms the hexapod structure and was thus chosen. The exact profile and dimensions of the struts still need to be determined.

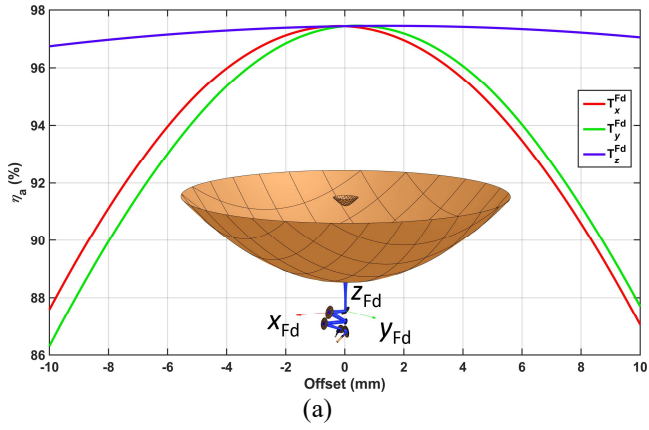

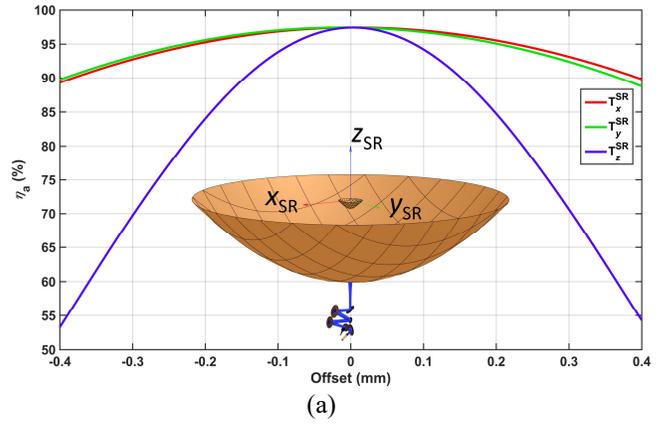

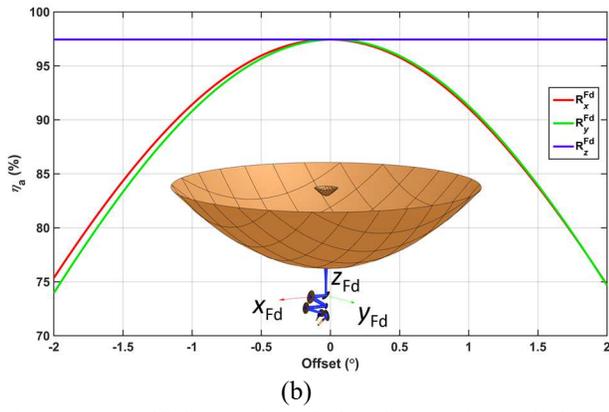

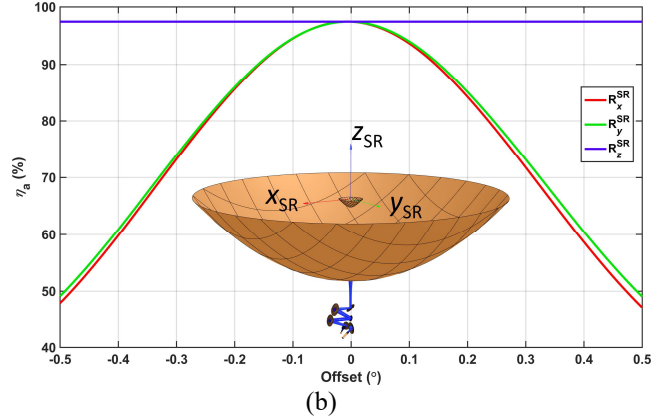

**Fig. 8.** Aperture efficiency degradation due to (a) translation on, and (b) rotation about, the principal axes of the coordinate system at the reflector system's focus of the downstream optics at 320 GHz.

**Fig. 9.** Aperture efficiency degradation due to (a) translation on, and (b) rotation about, the principal axes of the sub-reflector's coordinate system as shown in the inset at 320 GHz.

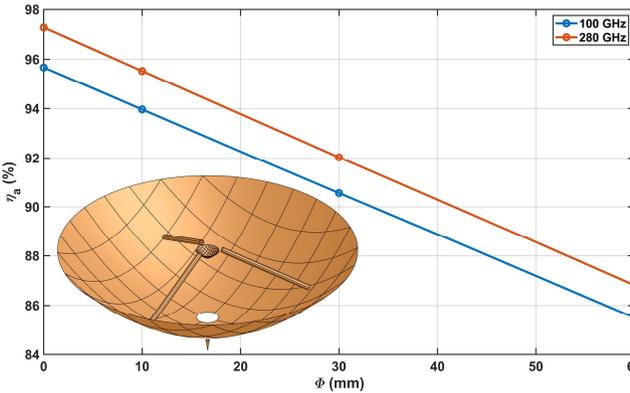

**Fig. 10.** Aperture efficiency degradation due to circular rod tripod strut configuration with various rod diameters.

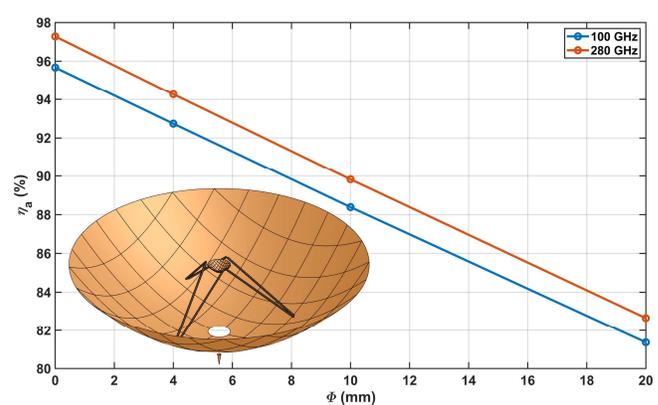

**Fig. 11.** Aperture efficiency degradation due to circular rod hexapod strut configuration with various rod diameters.

## V. Conclusion

We illustrated the preliminary baseline design of a shaped axially displaced Cassegrain (ADC) reflector system for the BHEX mission that maximizes the aperture efficiency. The considerations and trades resulting in this design and its optimization were presented. The efficiency achieved is above 95% for a perfect surface. The surface tolerance will degrade the performance by 20% (at the highest frequency) and the struts will add a maximum of another 10%.

## References

[1] M. Johnson, *et al.*, "The Black Hole Horizon Explorer: motivation and vision," in *Proc. SPIE Astronomical Telescopes + Instrumentation*, Yokohama, Japan, Jun. 2024.

[2] T.K. Sridharan, R. Lehmensiek, S. Schwarz, and D.P. Marrone, "Antenna technology readiness for the Black Hole Explorer (BHEX) mission," in *Proc. Int. Conf. Electromagn. Adv. Appl. (ICEAA)*, Palermo, Italy, Sep. 2025, to be published.

[3] D.P. Marrone, *et al.*, "The Black Hole Horizon Explorer: instrument system overview," in *Proc. SPIE Astronomical Telescopes + Instrumentation*, Yokohama, Japan, Jun. 2024.

[4] T.K. Sridharan, *et al.*, "The Black Hole Horizon Explorer: preliminary antenna design," in *Proc. SPIE Astronomical Telescopes + Instrumentation*, Yokohama, Japan, Jun. 2024.

[5] R. Lehmensiek, T.K. Sridharan, M. Johnson, and D.P. Marrone, "The Black Hole Explorer: mission overview and antenna concept," in *Proc. IEEE Int. Symp. AP & USNC/URSI Nat. Radio Sci. Meet.*, Florence, Italy, Jul. 2024.

[6] R. Lehmensiek and D.I.L. de Villiers, "An optimal 18-meter shaped offset Gregorian reflector for the ngVLA radio telescope," *IEEE Trans. Antennas Propag.*, vol. 69, no. 12, pp. 8282-8290, Dec. 2021.

[7] TICRA, TICRA Tools, 2024. [Online]. Available: https://www.ticra.com/ticratools/

[8] C. Granet, "A simple procedure for the design of classical displaced-axis dual-reflector antennas using a set of geometric parameters," *IEEE Antennas Propag. Mag.*, vol. 41, no. 6, pp. 64-71, Dec. 1999.

[9] F.J.S. Moreira and A. Prata, "Generalized classical axially symmetric dual-reflector antennas," *IEEE Trans. Antennas Propag.*, vol. 49, no. 4, pp. 547-554, Apr. 2001.

[10] F.J.S. Moreira and J.R. Bergman, "Shaping axis-symmetric dual-reflector antennas by combining conic sections," *IEEE Trans. Antennas Propag.*, vol. 59, no. 3, pp. 1042-1046, Mar. 2011.

[11] W.V.T. Rusch and R. Wohlleben, "Surface tolerance loss for dual-reflector antennas," *IEEE Trans. Antennas Propag.*, vol. 30, no. 4, pp. 784-785, Jul. 1982.

[12] C.E. Tong, K. Carter, P. Grimes, E. Lauria, D. Marrone, G. Montano, M. Morgan, Y. Uzawa, and L. Zeng, "Dual band receiver design for the Black Hole Explorer (BHEX) mission," in *Proc. Int. Symp. Space THz Technol. (ISSTT)*, Berlin, Germany, Apr. 2025.